\documentclass[fleqn,10pt]{wlscirep}
\usepackage[utf8]{inputenc}
\usepackage[T1]{fontenc}
\usepackage{bm}
\usepackage{braket}
\title{Higher-order band topology}

\author[1, 4, 6]{Biye Xie}
\author[2, 7]{Hai-Xiao Wang}
\author[1, 4]{Xiujuan Zhang}
\author[3]{Peng Zhan}
\author[2*]{Jian-Hua Jiang}
\author[1, 4, 5*]{Minghui Lu}
\author[1, 4*]{Yanfeng Chen}

\affil[1]{National Laboratory of Solid State Microstructures, Department of Materials Science and Engineering, Nanjing University, Nanjing 210093, China}
\affil[2]{School of Physical Science and Technologyand Collaborative Innovation Center of Suzhou Nano Science and Technology, Soochow University, 1 Shizi Street, Suzhou, 215006, China}
\affil[3]{School of Physics, Nanjing University, Nanjing 210093, China}
\affil[4]{Collaborative Innovation Center of Advanced Microstructures, Nanjing University, Nanjing 210093, China}
\affil[5]{Jiangsu Key Laboratory of Artificial Functional Materials, Nanjing 210093, China}
\affil[6]{Department of Physics and HKU-UCAS Joint Institute for Theoretical and Computational Physics at Hong Kong, 
The University of Hong Kong, Pokfulam Road, Hong Kong, China}
\affil[7]{School of Physical Science and Technology, Guangxi Normal University, Guilin, 541004, China}

\affil[*]{To whom correspondence should be addressed. e-mails: luminghui@nju.edu.cn, yfchen@nju.edu.cn, jianhuajiang@suda.edu.cn}

\begin{abstract}
The bulk-edge correspondence, i.e., a topological insulator (TI) has gapless edge states dictated by its bulk topological invariant, is commonly regarded as a distinctive signature of TIs. In the past years, a new type of TIs is found which hosts, instead, gapped edge states and gapless hinge or corner states protected by the bulk band topology. These unconventional TIs, termed higher-order TIs (HOTIs), provide counter-examples of the bulk-edge correspondence, are surprisingly common in crystalline and quasi-crystalline materials. Here, we review the principles, theories and materials of HOTIs for both electrons and classical waves. We put an emphasis on the developments of HOTIs in photonic, phononic and circuit systems due to their special contributions to the field. We elaborate on the physical mechanisms, experimental realizations, novel phenomena and potential applications in both electronic and classical systems. Based on these discussions, we remark on the trends and challenges in the field and the impact of higher-order band topology on other scientific disciplines. 
\end{abstract}
\begin{document}

\flushbottom
\maketitle

\thispagestyle{empty}

\section*{Introduction}

TIs with bulk-boundary correspondence (BBC)~\cite{topo1,topo2}, i.e., the emergent gapless edge states protected by the bulk band topology, provide a material platform with unprecedented electronic properties, such as back-scattering immune transport in the edge channels and Majorana modes with non-Abelian statistics in topological superconducting systems. TIs further fertilize the study of semimetals, yielding topological Weyl and Dirac semimetals with Fermi-arc surface states and exotic optoelectronic properties~\cite{RevModPhys.90.015001}. To date, TIs form a rich family of materials with various protecting symmetries, including the time-reversal symmetry, spatial symmetries (e.g., reflection and inversion symmetries) and crystalline symmetries. Although different protecting symmetries give rise to different topological invariants, TIs of different kinds all exhibit gapless edge states on the boundaries of the candidate materials.

When such a picture was anticipated to be true once for all, a new type of TIs, HOTIs, were proposed which have, instead, gapped edge states and gapless boundary states at lower dimensions~\cite{HOTI1,HOTI2,HOTI3,HOTI4,HOTI5,HOTI6}. For instance, a 2D HOTI has gapped edge states and in-gap corner states (Fig.~1a). A 3D HOTI may have gapped surface states and gapless hinge states (Fig.~1b), or gapped surface and hinge states but in-gap corner states (Fig.~1c). HOTIs broaden the scope of topological materials and provide a deeper understanding of band topology. The emergence of topologically-protected hinge and corner states reflects the intricate interplay between crystalline symmetry and band topology. In the pioneering works, it was shown that higher-order band topology can be described rigorously by new topological invariants~\cite{HOTI1,HOTI2,HOTI3,HOTI4,HOTI5,HOTI6}. These bulk topological invariants, together with the protective symmetries, dictate the emergence of the gapless lower-dimensional boundary states, which is recognized as the higher-order BBC. Therefore, HOTIs are not the consequence of fine-tunings of the boundary states, but are due to novel topological physics. In fact, the lower-dimensional topological boundary states (e.g., corner states and hinge states) in HOTIs cannot be removed by adding any perturbations on the surface if the protective symmetry is preserved. As illustrated in Ref.~\cite{HOTI6}, modifying the surface states by perturbations that respecting the bulk symmetry, the number of chiral hinge states do not change.

HOTIs offer access to rich band dispersions on the surfaces, hinges and corners which are valuable for materials and applications. For instance, topological hinge states can be used to form Majorana modes under superconducting proximity effects~\cite{braid1}. Such a system can over-perform the existing nanowire systems where multi-subband effects smear out the transport signature of the Majorana modes. In photonics, topological corner states provide a route towards robust cavity modes which are advantageous for quantum photonics, strong light-matter interaction and low-threshold topological lasing~\cite{HOTIlaser1,HOTIlaser2}.

In this Perspectives, we intend to provide a concise review on the achievements in the emerging field of higher-order topological physics and materials. Based on the current research progresses, we category HOTIs into three major groups: (1) HOTIs with corner states, (2) HOTIs with gapless hinge states, (3) HOTIs in noncrystalline systems and HOTIs in systems with synthetic dimensions. Based on these discussions, we then provide a summary and perspectives on the field. Specifically, we sequentially discuss some open challenges and future trends, the synergy and empowering other fields, potential applications of HOTIs and a brief summary.

\subsection*{The birth of HOTIs}

Historically, the emergence of HOTIs stems from several aspects. First, in studying the gapless boundary states described by the boundary Hamiltonian derived from the bulk Hamiltonian, one finds that they can be gapped out by extra symmetry breaking term such as adding a magnetic field perpendicular to the boundaries to break time-reversal symmetry. Hence, topological phase transitions of the boundary Hamiltonians are possible. As a consequence, at the interface of two topologically distinct boundary areas, lower-dimensional states such as zero-dimensional corner states in two-dimensional system and one-dimensional hinge states in the three-dimensional system emerge as the boundary states of the boundary Hamiltonian~\cite{FZ2013,FS2017,HOTI6,sitte2012}. Second, in previous studies of topological phases, the quantized topological quantities such as the charge polarization, the Hall conductanceand the magneto-electric polarizability that characterize conventional topological phases of matter can be expressed as dipole polarization in 1D, 2D and 3D systems respectively. One can then generalize this to higher-order polarizations such as the quadrupole and octupole polarization leading to novel topological phases with lower-dimensional boundary states such as corner states~\cite{HOTI1}. Besides, zero-dimensional topological states in a higher-dimensional systems can also be constructed by real space topological defects in topological crystalline insulators~\cite{non2}. Inspired by one dimensional Su-Schrieffer-Heeger (SSH) model, the higher-dimensional generalization of SSH model may host topological lower-dimensional boundary states which are protected by the crystalline symmetries. Moreover, by considering the zero eigenvalues in the local in-gap Green’s function, the lower-dimensional boundary states can emerge as higher-codimensional impurity-bound states~\cite{Slager2015}. Although there have been several sporadic works indicating the existence of HOTIs, the clear concept was not put forward until 2017~\cite{FS2017,HOTI6}. After that, an explosive number of works on HOTIs emerged and attracted tremendous research attention~\cite{HOTI2,HOTI3,HOTI4,HOTI5,HOTI7,HOTI8,HOTI9,HOTI10,HOTI11,HOTI12,HOTI13,HOTI14,HOTI15,HOTI16,HOTI17,HOTI18,HOTI19,HOTI20,HOTI21,HOTI22,HOTIBBC}.

Many works have been devoted to possible protective symmetries and the resultant higher-order band topology~\cite{HOTI3,HOTI5,HOTI6,HOTI11,HOTI15}. For instance, higher-order band topology can be protected by reflection symmetry~\cite{HOTI3}, rotation symmetry~\cite{HOTI5}, inversion symmtry~\cite{HOTI11} and combined time-reversal and rotation symmetry~\cite{HOTI6}. Here, we would rather give a simplified overview of the principles of HOTIs. Originating from topological crystalline insulators (TCIs)~\cite{PhysRevLett.106.106802}, HOTIs reflecting the phases where the gapless surface states are not protected by the crystalline symmetry. However, such phases cannot be transformed adiabatically into trivial atomic insulators, reflecting their topological nature. In many cases, HOTIs can be understood as special TCIs whose surface states are described by massive Dirac Hamiltonians. However, the signs of the Dirac masses are restricted by the crystalline symmetry. The sign reversal of the surface Dirac Hamiltonians for adjacent surfaces in a symmetry-preserving, finite-sized system leads to the emergent gapless hinge states. It has been demonstrated that such a picture already corresponds to a large classification of TCIs which have higher-order band topology. As can be perceived intuitively, the possibility of having gapped surface states and higher-order band topology is larger than the possibility of having gapless surface states in topological crystalline phases. Indeed, recent studies predict lots of material candidates for HOTIs (see, e.g., Ref.~\cite{wan}).

In the existing studies, photonic, phononic and other classical systems have shown superiority in designing, realizing and observing HOTIs. In fact, the material and experimental systems for such studies are much more versatile in classical systems than in electronic systems. For instance, using photonic and phononic metamaterials as well as electric circuits, various HOTIs have been realized. The topological corner and hinge states are then studied using local pump-probe measurements which are available in classical systems because of their macroscopic sizes or advantages in precise spectroscopy and local signal detection. To date, the experimental confirmation of the corner states in electronic systems is still absent. Topological hinge states in electronic materials are verified through scanning tunneling microscope and transport evidences. In contrast, in classical systems, HOTIs and the emergent topological corner and hinge states have been realized and observed in many different bsystems~\cite{HOTIEXP1,HOTIEXP2,HOTIEXP3,HOTIEXP4,3D1,3D2,3D3,3D4,HOTI3D1,HOTI3D2,HOTI3D3,HOTI3D4,HOTI3D5,non1,non3,non4,BIYEPRL,DJWPRL,optica,2019NP1,jiang1,lin,jiang2,jiang3,2019NP2,nano,2020NC,AS,Inter4,OL,AM,SB,elastic,lamb,circuit,spring,cold,quantum,johan1,onchip1,zhang1,Liew1,zhan2}. Furthermore, classical systems possess these additional advantages: (1) There is no Fermi level or band-filling effect in classical systems and therefore topological states can be measured at any energy/frequency directly. This advantage not only reduces the material complexity, but also enables the study of higher-order band topology in multiple band gaps in a single system. (2) The components and geometry of metamaterials for classical waves can be well controlled and thus give access to many possibilities in realizing various topological bands and even multiple sets of topological bands with all kinds of crystalline symmetries. (3) The band dispersions and the wave functions of the bulk and boundary states can be directly measured. Thus the study of topological phenomena can be verified straightforwardly in experiments. At this stage, in classical systems, higher-order topological phenomena are not only studied for fundamental science, but also for unprecedented material properties and their applications in photonics, phononics and other fields~\cite{2020LSA}.

In classical systems, the ability to manipulate waves at will is at the heart of many scientific and technological disciplines, such as photonics, acoustics and mechanics, sustaining vast applications from the state-of-art metamaterials to quantum, sensing, energy and communication technologies~\cite{photonics,phononics}. However, one of the main challenges in various disciplines is the existence of disorder-induced wave scattering  owing to unavoidable material imperfections. With the development of topological photonics~\cite{topopho1,topopho2,topopho3,topopho4}, topological phononics~\cite{topoaco1,topoaco2,topoaco3,topoaco4} and related fields, band topology and its higher-order generalization offer possible solutions to such challenges. The versatile topological boundary states in multiple dimensions in HOTIs provide a rich tool box to stabilize information and energy transport and manipulation and thus enables topologically-protected quantum/integrated photonics, topological lasing, topological switching and topological waveguiding that are valuable for science and technology frontiers. Because of the above reasons, a considerable portion of this paper will be devoted to classical systems.

\section*{HOTIs with corner states}

Topologically-protected corner states, which exhibit fully localized wavefunctions, signal the HOTIs in both 2D and 3D systems (Fig. 1a-b). In electronic materials, corner states are proposed to emerge in breathing kagome and pyrochlore lattices~\cite{kunst,HOTI7}, twisted bilayer graphene~\cite{HOTI10}, graphdiyne~\cite{PhysRevLett.123.256402} and the modified Kane-Mele model~\cite{niu}. In many cases, HOTIs also host fractional corner charges~\cite{HOTIC,frac1,HOTI10} which enable them to be identified in an alternative way. In higher-order topological superconductors~\cite{fan,braid1,nori,braid2,braid4,braid3,PhysRevLett.125.097001,PhysRevLett.125.107001,PhysRevLett.125.037001,PhysRevLett.123.167001}, the corner states provide a natural pathway toward localized Majorana zero modes~\cite{fan,braid3,nori} which can be instrumental for the non-Abelian braiding and topological quantum computation. We elaborate below several prototypes of HOTIs with corner states, while there are many other cases which cannot be covered here.

\subsection*{Quantized multipole insulators}

As a pioneering example of HOTIs, quantized multipole insulators (QMI), e.g., quadrupole and octupole topological insulators, was proposed in 2017. The defining property of the QMI is the absence of the bulk dipole polarization and the emergence of the quantized, fractional electric multipole moments in the bulk, as depicted in Fig. 2a~\cite{HOTI1}. A dimensional hierarchy of the boundary states and their multipole moments are in order: A quadrupole topological insulator has gapped edge states with quantized dipole polarization $P=\frac{1}{2}$ and in-gap corner states, accompanying with $\pm \frac{1}{2}$ fractional corner charges. An octupole insulator has gapped surface states with fractionally quantized quadrupole moment $q_{xy}=\frac{1}{2}$, gapped hinge states with quantized dipole polarization and in-gap corner states, accompanying with $\pm \frac{1}{2}$ fractional corner charges. The emergence of QMI requires the quantization of the dipole, quadrupole and octupole moments which needs reflection and inversion symmetries. So far, QMIs are constructed in square or cubic lattice systems. In addition, there must be at least two (four) bands below the band gap to cancel the total dipole (quadrupole) moment. Tight-binding models for quadruple (octupole) topological insulators were proposed in Refs.~\cite{HOTI1,HOTI2} using square (cubic) lattice models with four (eight) sites in each unit-cell for spinless electrons with coexisting positive and negative nearest-neighbor couplings, as shown in Figs.~2b and 2c. The negative and positive couplings enable the $\pi$-flux per plaquette, leading to two noncommutative reflection symmetries which is necessary for the emergence of the nontrivial quadrupole topology.

The quadrupole and octupole topological invariants can be calculated either using the nested Wilson loop approach or using the many-body multipole operators. For the nested Wilson loop approach, one may first construct the Wilson loop operator along the $x$ and $y$ directions, $W_{x, {\bf k}}$ and $W_{y, {\bf k}}$, with the starting point of the loop being ${\bf k}=(k_x, k_y)$. The eigenvalues and eigenstates of the Wilson loop operators, $W_{x, {\bf k}}$ and $W_{y, {\bf k}}$, form the Wannier bands and the associated Wannier wavefunctions. By constructing the Wilson loops for the Wannier bands to characterize the quantized polarization of the Wannier bands below the Wannier gap, one can obtain the quadrupole moment of the bulk bands (see Figs. 2c and 2d). The octupole moment can be calculated using the Wilson loop of the nested Wannier bands (see Figs. 2e, 2f and 2g). A more intuitive approach to calculate the multipole topological invariants is to use the so-called many-body multipole operators proposed in Refs.~\cite{manybody1,manybody2}. These two approaches have their own advantages and disadvantages.

There are higher-order topological insulators and semimetals derived from the quantized multipole insulators. For instance, in 3D systems, 3D quadrupole semimetals emerge as an intermediate phase between the trivial insulator and the 3D quadrupole insulator. Recently, higher-order Weyl semimetals are proposed as the intermediate phase between the 3D quadrupole insulator and the normal Weyl semimetal in systems with four-fold rotation or screw symmetry~\cite{howp1,howp2}. Such a higher-order Weyl semimetal is realized recently in sonic crystals~\cite{howp3}.

\subsection*{Experimental realizations}
Soon after the establishment of the theory, experimental realization of quadrupole topological insulators are reported~\cite{HOTI1,HOTI2}. The key challenge is to realize a tight-binding model (TBM) with coexisting positive and negative nearest-neighbour couplings which enable a $\pi$-flux square lattice. This challenge is solved using mechanical metamaterials and microwave transmission-line metamaterials. In perturbative mechanical metamaterials the coupling between local resonances can be engineered by the mechanical connection between them. A nondegenerate resonance with its frequency separated from other resonances are preferred. The connections between regions of the same sign (opposite signs) of the resonance wavefunction realize the positive (negative) couplings~\cite{HOTIEXP1}. Similar design principle is exploited in the microwave transmission-line metamaterials to realize the coexisting positive and negative couplings~\cite{HOTIEXP2}. In phononic resonator network, the negative inter-resonator couplings are achieved
by connecting the resonators with thin waveguides on different
sides of each resonance’s nodal line. Positive and negative couplings can also be easily realized in electric-circuits by simply adding inductors and capacitors between lattice sites. Using such a design principle, quadrupole topological insulators are first realized in mechanical and transmission-line metamaterials and then in electric circuits~\cite{HOTIEXP3}, coupled optical ring resonators~\cite{2019NP1}, sonic crystals~\cite{HOTIEXP4} and plasmon-polaritonic systems~\cite{jiang3,Liew1}.

Thanks to the diverse and controllable geometry and structure of classical systems, it is also possible to realize quadrupole topological insulators without negative couplings. As shown in sonic crystals~\cite{lin,jiang2} and photonic crystals~\cite{jiang3}, one can achieve quantized quadrupole moments in nonsymmorphic $p4g$ lattices. In such nonsymmorphic crystals, the quadrupole topology is protected by two noncommutative glide symmetries. Besides, another group reported that in 2D photonic crystals consisting of gyromagnetic materials (e.g., yttrium iron garnet), if a magnetic field is applied to break the time-reversal symmetry, the quadrupole moment is quantized by the combined time-reversal---mirror (although time-reversal or mirror alone is not symmetry operation) symmetries~\cite{2020NC}. Realizing 3D octupole topological insulators is much harder than the realization of 2D quadrupole topological insulators since achieving a full band gap in classical periodic systems is more challenging in 3D than in 2D. Nevertheless, it was reported that by using a 3D network of coupled acoustic resonators~\cite{3D1,3D2} and LC resonators in electric circuits~\cite{3D3,3D4}, one can directly mimicking the 3D tight-binding model of octupole topological insulators. Through this approach, octupole topological insulators with gapped surface and hinge states and eight in-gap corner states are observed.

\subsection*{HOTIs based on generalized SSH models}
The QMIs, which require noncommutative mirror symmetries, are often difficult to realize in materials. A natural question arise is whether there are HOTIs that do not require such conditions and can be realized in a more straightforward way? The answer is yes. In the literature, HOTIs without quantized multipole moment were first proposed theoretically~\cite{HOTI7,HOTIC} and then realized in various experimental systems. They are relatively easier to realize and have various advantages in materials and applications (see Table 1).

The original SSH model describes 1D spinless electronic systems in a chain of orbits with alternative strong and weak couplings~\cite{SSH1}. In the topological phase, which is characterized by the Zak phase $\pi$ or equivalently the fractional bulk dipole polarization (see Fig. 3a), there are in-gap edge states appearing at the ends of a finite chain. In addition, fractional charges $\pm e/2$ emerge at the two ends. With chiral symmetry, the in-gap edge states are signatures of the bulk band topology.

The 1D SSH model can be generalized to higher dimensions. A prototype of HOTIs without quantized multipole moment is the 2D SSH model in square-lattice materials (see Fig. 3b-c~\cite{SSH2,SSH3,non1}). By changing the relative strength between intercell and intracell couplings, the system can experience a topological phase transition. If the intercell coupling is stronger than the intracell coupling, the system has a fractional, quantized bulk dipole polarization ${\bf P}=(\frac{1}{2},\frac{1}{2})$. At the filling factor $\frac{1}{4}$ (i.e., filling only the lowest band), the Wannier center locates at the corner of the unit-cell which indicate a nontrivial phase with fractional corner charge of $\frac{1}{4}$ (see Fig. 3a). Accompanying with the fractional corner charge, the corner states emerge at the corner boundaries (Fig. 2d-2e). Here, the higher-order topology is induced by the filling anomaly: the mismatch between the number of electrons required to simultaneously satisfy charge neutrality and the crystalline symmetry leads to a fractional corner charge of $1/4$. Based on the symmetry representations of the occupied energy bands, a filling anomaly topological index is constructed which can quantitatively yield the fractional corner charge (see Fig. 3a). If the $C_4$ rotation symmetry is reduced to the $C_2$ symmetry, one can  obtain other phases. For instance, the insulating phases with the bulk dipole polarization ${\bf P}=(\frac{1}{2},0)$ or $(0,\frac{1}{2})$. These phases do not support corner states but support edge states in the $x$ or $y$ direction as induced by the nontrivial Zak phase in the perpendicular direction~\cite{SSH2}. We note that HOTIs based on generalized SSH models is intrinsically different from QMIs in the following aspects: (1) QMIs are characterized quadrupole or octupole topological index due to the gapped Wannier bands, whereas the generalized SSH models do not have such properties; (2) QMIs have vanishing dipole polarizations, whereas generalized SSH models may not have; (3) QMIs require noncommutative reflection symmetries, whereas the generalized SSH models do not require such conditions. In short, these two kinds of HOTIs have different topological indices, protective symmetries and mechanisms.

The first proposed realization of the 2D SSH model in classical waves was achieved in all-dielectric photonic crystals in 2018 using alumina rods in a similar lattice configuration as the 2D SSH model (see the middle panel of Fig. 3a)~\cite{non1}. By tuning the geometry parameters $l_x$ and $l_y$, the authors find a topological phase diagram characterized by the bulk dipole polarization ${\bf P}$. When ${\bf P}=(\frac{1}{2},\frac{1}{2})$, the system is in the HOTI phase which supports coexisting edge and corner states for finite systems. The experimental observation of such a HOTI is achieved in Refs.~\cite{BIYEPRL,DJWPRL} for microwave photonic crystals as shown in the right panel of Fig.~3a. Due to the scale invariance of the Maxwell's equations, the design in Ref.~\cite{non1} can be directly generalized to optical frequencies using sub-micron dielectric structures. Later, this photonic HOTI was realized in near-infrared photonic crystals~\cite{optica} and holey sonic crystals~\cite{AM}. Plasmonic nanoparticle arrays are also proposed to host such a 2D HOTI phase and demonstrate polarization-dependent corner states~\cite{nano}.

The SSH model can be further generalized to other 2D lattices, e.g., 2D lattices with $C_6$ or $C_3$ symmetry. The $C_6$ symmetric model with dimerized intercell and intracell couplings is in fact the first model that has been realized in real materials~\cite{non2} (see the left panel of Fig.~3b). In fact, before the knowledge of HOTIs, this model is regarded as a pathway to simulate the quantum spin-Hall-like effect in photonic and acoustic systems~\cite{SCIREP2014,spin3,spin5,spin9}. Recently, these topological states are revealed as HOTIs with coexisting edge and corner states~\cite{HOTISPIN2,HOTISPIN3,HOTISPIN4}. When the 2D SSH model is generalized to $C_3$ symmetric systems, this can yield HOTIs which simulate quantum valley-Hall effects in classical systems. A typical example is the breathing kagome HOTI which demonstrates corner states and gapped edge states in a triangular supercell~\cite{HOTI7}. The kagome HOTI is realized in sonic crystals~\cite{non3,non4} (see the right panel of Fig.~3b), coupled optical fibers for visible frequency light~\cite{2019NP2}, photonic crystals with long-range interactions~\cite{Inter} and electric-circuits~\cite{Ezawae}. For $C_2$ cases, HOTIs were proposed to appear in all-dielectric photonic crystals~\cite{BIYEPRL} and were observed in waveguide arrays with single-photon dynamics~\cite{quantum}.

The generalization of the SSH model to 3D systems has been achieved in the literature. Theoretically, it was proposed that 3D polyhedra lattice~\cite{HOTI7} can host a third-order topological insulating phase with gapped surface and hinge states and in-gap corner states. This phase was later observed in acoustic resonators arrays~\cite{HOTI3D1,HOTI3D2} (see Fig.~3c). The 3D SSH model in cubic lattices has been realized in sonic crystals where a third-order topological insulating phase is observed~\cite{HOTI3D3}. Since the space group symmetry is much larger in 3D than in 2D, there are much more possibilities for 3D models and systems to host HOTI phase in the manner of generalized SSH models~\cite{HOTI3D4,HOTI3D5}.

\subsection*{Multidimensional topological physics and topological transitions in HOTIs.}

HOTIs are unique not only because of lower-dimensional topological boundary states, but because it has an integration of multidimensional topological physics induced intricately by the crystalline symmetry~\cite{HOTI1,HOTI2}. For an $n$D HOTI with $n>1$, all the boundary states at $(n-1), (n-2), ..., 1$D sub-systems are induced by topological mechanismsand they themselves can be regarded as topological insulators at lower dimensions. Such an integration of topological mechanisms at different dimensions is a key feature of HOTIs.

Experimentally, 2D sonic crystals with configurable geometry were used to visualize such multidimensional topological physics~\cite{zhang1}. The 2D acoustic bands are revealed as analogs of 2D massive Dirac systems where the Dirac mass can be controlled by the geometry of the sonic crystal~\cite{zhan2}. Such an analog leads to the 1D edge states emerging at the edge boundaries between the sonic crystals with Dirac masses of opposite signs. Furthermore, the 1D edge states are gapped and have quantized Zak phases, i.e., they can be 1D topological insulators. These edge states can also be regarded as 1D massive Dirac systems. The sign change in the 1D Dirac masses of the $x$- and $y$-edges leads to the emergence of the 0D corner states. This chain of logic reveals topological mechanisms in a hierarchy of dimensions.

Another important feature of HOTIs is the multidimensional topological transitions which were discovered in Ref.~\cite{zhang1}. For such a purpose, topological transitions of the edge states without closing the bulk band gap as controlled by a geometry parameter are exploited. Through such a geometric control, both the edge and bulk topological transitions can be triggered independently. Using these multidimensional topological transitions, it is demonstrated that rich scenarios for the transfer between 0D corner states, 1D edge states and 2D bulk states can be realized. Multidimensional topological transitions were later found in other systems~\cite{HOTI8,jiang2,lin}.

Topological boundary states at multiple dimensions have been observed in various classical systems~\cite{zhang1,zhan2,HOTI3D1,HOTI3D2,HOTI3D3,HOTI3D4,HOTI3D5}. They have been regarded as a key experimental signature of HOTIs. For instance, in 2D photoinc SSH model, both 1D edge states and 0D corner states has been observed~\cite{non1}. In 3D HOTIs, both 2D surface states, 1D hinge states and 0D hinge states has been shown in 3D acoustic systems~\cite{HOTI3D1,HOTI3D2,HOTI3D3,HOTI3D4,HOTI3D5}.

In applications, the multidimensional topological physics and topological transitions could support multifunctional and integrated topological devices. For instance, in integrated photonics (the photonic analog of integrated circuits) the two fundamental elements, waveguides and cavities, can be realized using HOTIs which support coexisting edge and corner states (functioning as waveguides and cavities, respectively). The coupling between the 1D edge states and the 0D corner states in linear, nonlinear and non-Hermitian systems enable more functions~\cite{jiang1,Liew1}. Such capability of multidimensional topological wave trapping, guiding and manipulation is absent in conventional photonic TIs. Furthermore, controllable multidimensional topological phase transitions may enable topological transfer~\cite{zhang1} and topological switches~\cite{HOTI8}.

\section*{HOTIs with gapless hinge states}

Beside the aforementioned 0D topological corner states, 3D HOTIs may also host 1D gapless hinge states which are enforced by bulk crystalline topological invariants (such as the mirror Chern number) and protected by more general symmetries (such as the spatiotemporal symmetry $\mathrm{C_4T}$), mediating a spectral flow between valence and conduction band of the bulk insulator~\cite{HOTI6}. In electronic  materials, gapless hinge states are expected to be observed more easily than corner states as they can appear more generically in crystalline structures~\cite{HOTI14}. In spin $1/2$ HOTIs, regarding the existence of the time-reversal symmetry and the propagations of states, HOTIs with gapless hinge states can be classified into two major categories:(i) chiral hinge states with hinge states propagating unidirectionally akin to the surface quantum anomalous Hall effect, denoted as chiral HOTIs (see Fig. 1c left panel); (ii) helical hinge states with Kramers pairs of counterpropagating  hinge states resembling the 2D quantum spin Hall effect denoted as helical HOTIs~\cite{HOTI6}(see Fig. 1c right panel).

Theoretically, topological chiral hinge states have been studied in $\mathrm{C_4T}$ symmetric 3D HOTIs with broken time-reversal symmetry~\cite{HOTI6}, 3D quadrupole topological insulator with quantum pumping~\cite{HOTI2}, layer-stacking of inversion-symmetric quantum Hall layers~\cite{HOTI11,inversion1}, 3D strong topological insulators in magnetic fields~\cite{sitte2012}, reflection-symmetric 3D HOTIs~\cite{HOTI3}, 3D magnetic second-order topological insulators~\cite{ezawahinge}, axion topological insulators~\cite{axion3} and rotoinversion symmetric systems~\cite{HOTI15}. It was revealed that the in-gap circular dichroism, a spatially averaged, frequency integrated quantity is quantized and can differentiate different types of chiral HOTIs~\cite{pozo}. In real materials, the chiral hinge states was predicted to exist in 3D topological insulators with noncollinear antiferromagnetic order at low temperatures~\cite{HOTI6} and in the axion topological insulator $\mathrm{EuIn_2As_2}$~\cite{axion1}. Lately, Experiments show that the axion topological insulator $\mathrm{Bi_{2-x}Sm_xSe_3}$ is a chiral HOTI~\cite{axion2}. Considering the interaction between electrons, a 3D fractional HOTI with chiral hinge states has been proposed~\cite{fractional}.

Different form the chiral hinge states, helical hinge states are protected by the time-reversal symmetry and other spatial symmetries~\cite{HOTI5,HOTI6}. They can arise in a system with both $\mathrm{C_4}$ and $\mathrm{T}$ symmetries or with both mirror symmetries and $\mathrm{T}$~\cite{HOTI6}. First-principle calculations show that $\mathrm{Sm}$-doped $\mathrm{Bi_2Se_3}$~\cite{HOTI6}and transition metal dichalcogenides $\mathrm{XTe_2}$ $\mathrm{(X =Mo;W)}$~\cite{wan,bernevig} may be the candidates for helical HOTIs. By applying scanning-tunnelling spectroscopy and Josephson interferometry, bismuth was experimentally confirmed to be the first realistic helical HOTI~\cite{HOTI14}. Lately, an anisotropic helical hinge states have been spatially resolved by analysing the magnetic field interference of the supercurrent in $\mathrm{Nb-WTe_2-Nb}$ proximity Josephson junctions~\cite{anisotropic,PhysRevLett.124.156601}. Recently, higher-order topological superconductors with helical hinge states have been demonstrated in $\mathrm{Fe}$-based superconductor $\mathrm{FeTe_{0.55}Se_{0.45}}$~\cite{nanolett} and other systems~\cite{PhysRevLett.124.046801,sarma,HOTI11,PhysRevLett.123.167001}. There are also studies on HOTIs with gapless helical hinge states and surface rotation anomaly~\cite{FangFu}.

\section*{HOTIs in other systems}

\subsection*{HOTIs with synthetic dimensions}

Characterized by codimensions of the gapless boundary states, HOTIs are closely related to the dimensions of systems. Limited by spatial dimensions of realistic materials which are only in 0D, 1D, 2D, 3D systems, synthetic dimensions~\cite{syn} become an important tool to explore various HOTIs such as octupole insulator in a 3D cubic lattice and a hexadecapole (16-pole) insulator in a 4D hypercubic lattice~\cite{HOTIsyn1}. The principle of realizing a synthetic dimensional system is to couple internal  degrees of freedom such as the internal atomic states, orbital angular momentum, spatial eigenmodes, etc, to serve as extra spatial dimensions. Actually, by using the frequency as a dimension, one can realize a 2D quadrupole HOTI in a 1D array of modulated photonic ring resonators~\cite{HOTIsyn1}. If the frequency and orbital angular momentum of light are regarded as 2D synthetic space, a quadrupole topological insulators can be realized in 0D optical cavity~\cite{HOTIsyn2}. Synthetic dimensions can also be used to realized HOTIs in higher-dimensions\cite{HOTIsyn3,HOTIsyn4}. Beside being a platform for studying unconventional HOTIs, the synthetic dimensions also support the exploration of interacting HOTIs and corresponding many-body physics in classical waves. This is because compared with HOTIs in spatial dimensions where particles mostly interact locally, in HOTIs with synthetic dimension, the particles may interact with each other even though they are far away in parameter space as they can still be spatially close to each other~\cite{syn}. HOTIs in synthetic dimensions with many-body interactions open a frontier which is yet to be explored in both theory and experiments which may lead to novel quantum phases in interacting bosonic systems.

\subsection*{HOTIs in quasicrystals and amorphous lattices}

Rooted in the energy band theory, HOTIs for classical waves are not restricted to periodic systems. As band structures, energy gaps and topological phases can emerge at non-periodic lattices such as in quasicrystals, amorphous lattices and fractals\cite{HOTIamor1}, HOTIs may also be realized in these systems. Theoretically, HOTIs have been proposed in amorphous solid~\cite{HOTIamor2}, the quasicrystalline lattices with different tiling patterns~\cite{HOTIamor3}, the Kane-Mele model with a $30$ degree twist~\cite{HOTIamor4}and a superconductor on the Ammann-Beenker tiling~\cite{HOTIamor5} and a dodecagonal quasicrystal~\cite{HOTIamor6}. Due to the complex coupling configurations, so far there is no experimental realization of HOTIs in non-periodic systems which may be achieved in electric circuits and acoustic resonant networks.

\section*{Perspectives and a summary}

\subsection*{Open challenges and future trends}

Although HOTIs can provide coexisting multidimensional topological boundary states, the topological corner states are often not fixed to the middle of the band gap in classical systems~\cite{BIYEPRL}. This is essentially due to the lack of the chiral symmetry or particle-hole symmetry. In 2D SSH tight-binding model with only nearest-neighbour couplings, the chiral symmetry is preserved and the corner states are pinned at zero energy. However, in classical systems, except for some special systems directly mimicking the tight-binding model, the condition for only having nearest-neighbour coupling is always violated. For example, in 2D all-dielectric photonic crystals~\cite{BIYEPRL,DJWPRL}, the first band is beyond the description of tight-binding models with nearest-neighbour coupling. As a consequence, the chiral symmetry is naturally broken and the corner states in 2D photonic crystals is separated from the bulk states and edge states. In comparison, in the 2D SSH model with only the nearest-neighbor couplings, corner states and edge states are all pinned at zero energy and thus mixed together. In 3D acoustic systems, the corner states, hinge states and surface states can either mixed with the bulk states or well separated from each other in the band gap, depending on whether there are long-range couplings~\cite{HOTI3D1,HOTI3D2,HOTI3D3}. 
 
Without the chiral symmetry or particle-hole symmetry, the boundary states can either be in the gap or embedded into the bulk spectrum. The latter case is called as the higher-order bound states in continuum~\cite{bound1,bound2} and there is even no gapless boundary states within the bandgap and thus cannot be distinguished from trivial insulators simply by investigating the spectrum alone. To solve this problem, recently Peterson {\sl et al.} demonstrated that by measuring the mode densities of the bulk, edge and corner states, a fractional corner anomaly~\cite{HOTIC} can be quantitatively measured to reveal higher-order topological phases in 2D transmission line systems~\cite{frac1,defect2,frac2}.

In previous experimental realizations in classical waves, HOTIs are regarded as spinless models. Recently, spinful HOTIs have been theoretically investigated in tight-binding models~\cite{HOTISPIN1} and experimentally demonstrated in photonic crystals~\cite{HOTISPIN2}, sonic crystals~\cite{HOTISPIN3,HOTISPIN4}, mechanics~\cite{johan1,onchip1}. Intriguingly, when spin-momentum locking is considered, a higher-order quantum spin Hall effect is realized which extends the conventional quantum spin Hall effect into the higher-order version. So far, experimental realizations of 3D spinful HOTIs are still very limited, especially in classical systems. We expect to see a rapid development of realizing various spinful 3D HOTIs in classical wavesand in quick succession, designing novel 3D topological devices base on such spinful higher-order topological effects.

\subsection*{Synergy and empowering other fields}

As a fundamental research field, HOTIs can be synergized with other fields which will bring new physical effects, stimulating and empowering the developments of other fields. Below we list some possibilities.

First, the concept of HOTIs can be directly generalized to superconductors leading to higher-order topological superconductors with Majorana modes at corners and hinges~\cite{HOTI4,HOTI11,HOTI12}. Higher-order band topology in superconductors thus provides more options for the realization and manipulation of Majorana fermions which are believed to be the building block of the fault-tolerant topological quantum computing.

Second, HOTIs can be realized in the non-Hermitian regime to yield new physics and effects. Recently, non-Hermitian topology has been studied to reveal many ubiquitous physical properties such as the non-Hermitian skin effect, the non-Bloch bulk-boundary correspondence and non-Hermitian topological classifications~\cite{non-her1,non-her2}. In classical waves, the non-Hermitian regime can be induced by the gain and loss in photonics and phononics, the friction in mechanics, the dissipative heating in electric circuits, etc~\cite{non-her3}. When non-Hermitian effects are considered in HOTIs, many new physical effects emerge such as the higher-order skin effect, the biorthogonal bulk-boundary correspondence, the gain and loss induced HOTIs,~\cite{nh1,nh2,nh3,nh5,nh6,nh7,nh8,nh9}. This effect was proposed to be implemented in sonic crystals~\cite{nh10} and electric circuits~\cite{nh11}. Recently, a higher-order TI induced by deliberately introduced losses is experimentally realized in a phononic crystal~\cite{ZBLnon}.

Third, topological phases are not restricted to static systems~\cite{floq1,floq2}. A periodic time modulation induced by circularly-polarized irradiation or an
alternating Zeeman field can turn a trivial band insulator into a topological insulator, denoted as the Floquet topological insulators. This Floquet engineering offers more controllability and loose the strict requirement on material structures. Floquet HOTIs have been proposed by considering multistep driving of topologically trivial Hamiltonians, exhibiting Floquet quadrupole and octupole insulators with zero and/or $\pi$ corner modes~\cite{flo1}. They can also be constructed from an approximate time-glide symmetry~\cite{flo2}, anomalous dynamical polarization~\cite{flo3}and other periodic driving protocols~\cite{flo4,flo5,flo6,flo7}. In classical waves, the time modulations can be mimicked by a modulation in an extra spatial dimension such as the longitudinal direction of the laser-writing optical
waveguides~\cite{flowave}, 2D arrays of ring resonators supporting spoof plasmons~\cite{floreson}and coupling networks in a generalized spatial Floquet acoustic lattice~\cite{floaco}. Therefore, although not yet been realized, in principle, it is possible to implement various Floquet HOTIs in classical waves.

Although being focused on classical waves, one can also study quantum optics in photonic HOTIs~\cite{quan1,quan2,quan3}. The protection of the quantum correlation and entanglement is crucial for quantum communication and computation which is vulnerable to defects in materials~\cite{quan4}. As shown in Wang {\sl et al.} the corner states in a 2D photonic HOTI can promote the protection of the quantum correlation and entanglement\cite{quantum}. Furthermore, when combined with quantum emitters, one can achieve a chiral quantum emission of light by using HOTIs~\cite{quan5}.

Another promising area is to realize HOTIs by using metamaterials (i.e., artificial composite materials with physical properties not easily found in natural materials)~\cite{meta0}. For instance, compared with photonic crystals, photonic metamaterials offer accessibility to arbitrary designs of the permittivity and permeability tensors which will broaden the scope of physics and materials for HOTIs. The metamaterials have been successfully used to realize 3D photonic topological insulators and various other topological states.  Recently, it was proposed that HOTIs can be realized in metasurfaces~\cite{meta1,meta2,meta3}. Realizing and studying HOTIs in various metamaterials will be an emergent trend.

Combining HOTIs with nonlinear physics will be a very interesting topic, especially in classical waves. The nonlinearity can be introduced in many classical systems such as the optical waveguides, resonators, atomic gases and metamaterials~\cite{nonlinear1}. Intriguingly, the nonlinear topological classical waves systems can enable many functional devices with nonreciprocity, active tunability, frequency conversion and many-body interactions~\cite{inter1,inter2}. Recently, nonlinear higher-order topological insulators has been proposed in a exciton–polariton system realized with a kagome arrangement of
microcavity pillars~\cite{nonlinear2} and demonstrated in electric circuits which enable dynamic tuning of the spectral properties and localization of the topological edge and corner states~\cite{nonlinear3}. This may lay the foundation for reconfigurable topological devices.

Finally, photonic HOTIs can interact with 2D materials such as transition metal dichalcogenides~\cite{TMD}, graphene~\cite{graphene}, perovskites~\cite{P}, etc, to enable unprecedented optoelectronic systems for topological enhancement and control of light emission, lasing as well as nonlinear and quantum optics~\cite{2DM1,2DM2,2DM3}. For example, one may combine a Mo$\mathrm{Se}_2$ with GaP-based slab photonic crystal to control the spontaneous emission~\cite{MOS2} which may significantly reduce the lasing threshold~\cite{2Dlaser}. HOTIs which possess diverse types of localized states may serve as the source  of emission and enable many tunable nano-photonic devices.

\subsection*{Potential applications}

In classical systems, HOTIs provide unprecedented topological wave trapping and manipulation at boundaries of different dimensions which is advantageous for various applications. Even at this stage, several potential applications of HOTIs have been demonstrated in classical systems. For instance, by introducing gain and loss, one can generalize HOTIs into topological lasers with higher-order band topology~\cite{HOTIlaser1,HOTIlaser2}. The topological corner states are excellent candidates for optical cavity modes. It was shown that in photonic HOTIs based on the 2D SSH model, the topological corner states emerge as high quality factor cavity modes in the photonic band gap~\cite{optica,Q}. The strong confinement of the wavefunctions of the photonic corner states enhances the energy intensity of photons which is then instrumental for realizing nano-lasers with low-threshhold~\cite{HOTIlaser1,HOTIlaser2}. In addition, in classical waves such as in photonics, phononics and electric circuits, real-space topological defects enable the realization of Majorana-like bound states and non-Abelian braiding statistics~\cite{braid1,braid2,braid3,braid4,braid5,Ezawab}. It was also found that the topological corner states also possess non-Abelian braiding properties~\cite{braid6,braid7}.

In 3D HOTIs, the dispersive chiral and helical hinge states in spinful systems are crucial for unidirectional transportations of classical waves in lower-dimensions. In electronic materials, hinge states are also interesting as they provide lossless electronic transport owing to their local protection from backscattering by disorders. When proximitized with superconductivity, hinge states in a nanowire with hexagonal cross-section enables a convenient way to build a hexon which has been proposed as building blocks for a measurement-only quantum computer~\cite{HOTI6,hexon}.

\subsection*{Summary}

In summary, we have concisely introduced the burgeoning field of higher-order band topology. As revealed by the existing studies, HOTIs widely exist in various electronic and classical systems and can bring about unprecedented lower-dimensional robust boundary states and multidimensional topological physics. As elaborated above, the concept of HOTIs further generates other physical concepts. For instance, higher-order Weyl semimetals were proposed and realized lately~\cite{howp1,howp2,howp3,howp4}). Recently, novel experimental methods that characterizes the fractional charge at the corner boundaries~\cite{frac1} or at the disclination cores~\cite{defect2,frac2} provide unprecedented approaches to identify and distinguish various HOTIs in experiments. To illustrate the broad influence of HOTIs, we list some promising topics at the boundaries between HOTIs and other research fields. These topics will bring new physical effects, stimulating and empowering the development of many other scientific areas. Perspectives of HOTIs on future applications are depicted and summarized in Fig.~4. Looking forward, the study of HOTIs in classical systems is a new horizon with many interesting and interdisciplinary aspects, including the underlying physics, phenomena, materials and applications that are still waiting for further explorations and enrichments.





\noindent

\bibliography{Ref_NRPv1}

\section*{Acknowledgements}
This work was supported by the National Key
R \& D Program of China (Grants No. 2017YFA0303702and No. 2018YFA0306200, No. 2016YFB0700301), 
the National Natural Science Foundation of China (Grants
No. 11625418, No. 11474158, No. 11890700, No. 51732006, No. 11904060, and No. 12074281) and the Jiangsu specially-appointed professor fundings.

\section*{Author contributions}
All authors worked together on preparing and writing this Perspective.

\section*{Competing interests}
The authors declare no competing interests. 

\newpage

\begin{figure}[ht]
\centering
\includegraphics[width=\linewidth]{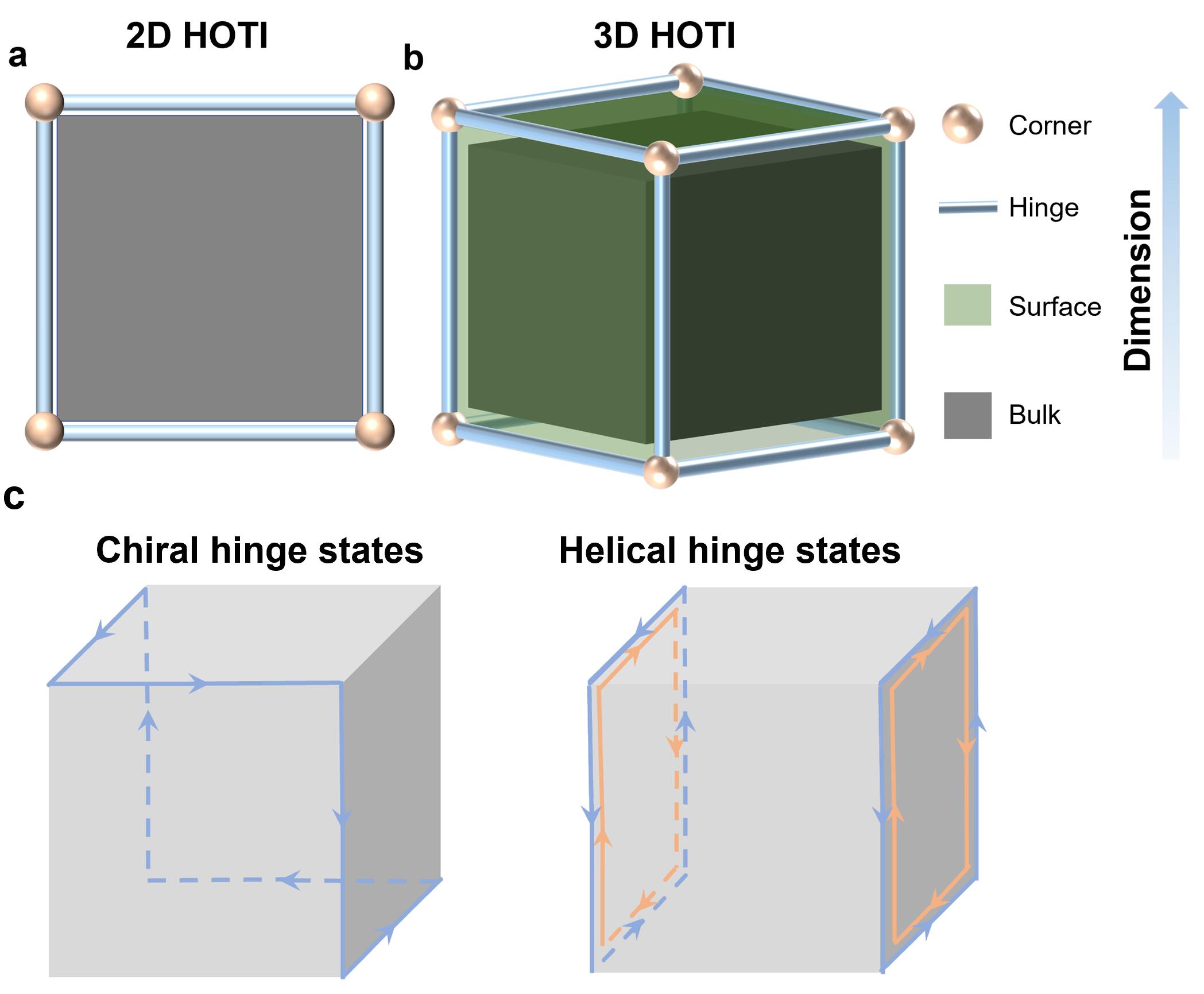}
\caption{\textbf{Higher-order topological insulators (HOTIs) (denoted as black areas) with multidimensional topological boundary states and their classification.} \textbf{a} For 2D HOTIs, there are 1D topological hinge states (denoted as blue lines) and 0D topological corner states (denoted as pink spheres). \textbf{b} For 3D HOTIs, there are 2D topological surface states (denoted as green surfaces), 1D topological hinge statesand 0D topological corner states. \textbf{c} Schematic of chiral hinge states and helical hinge states in 3D HOTIs~\cite{HOTI3,sarma}. }
\label{fig}
\end{figure}

\begin{figure}[ht]
\centering
\includegraphics[width=\linewidth]{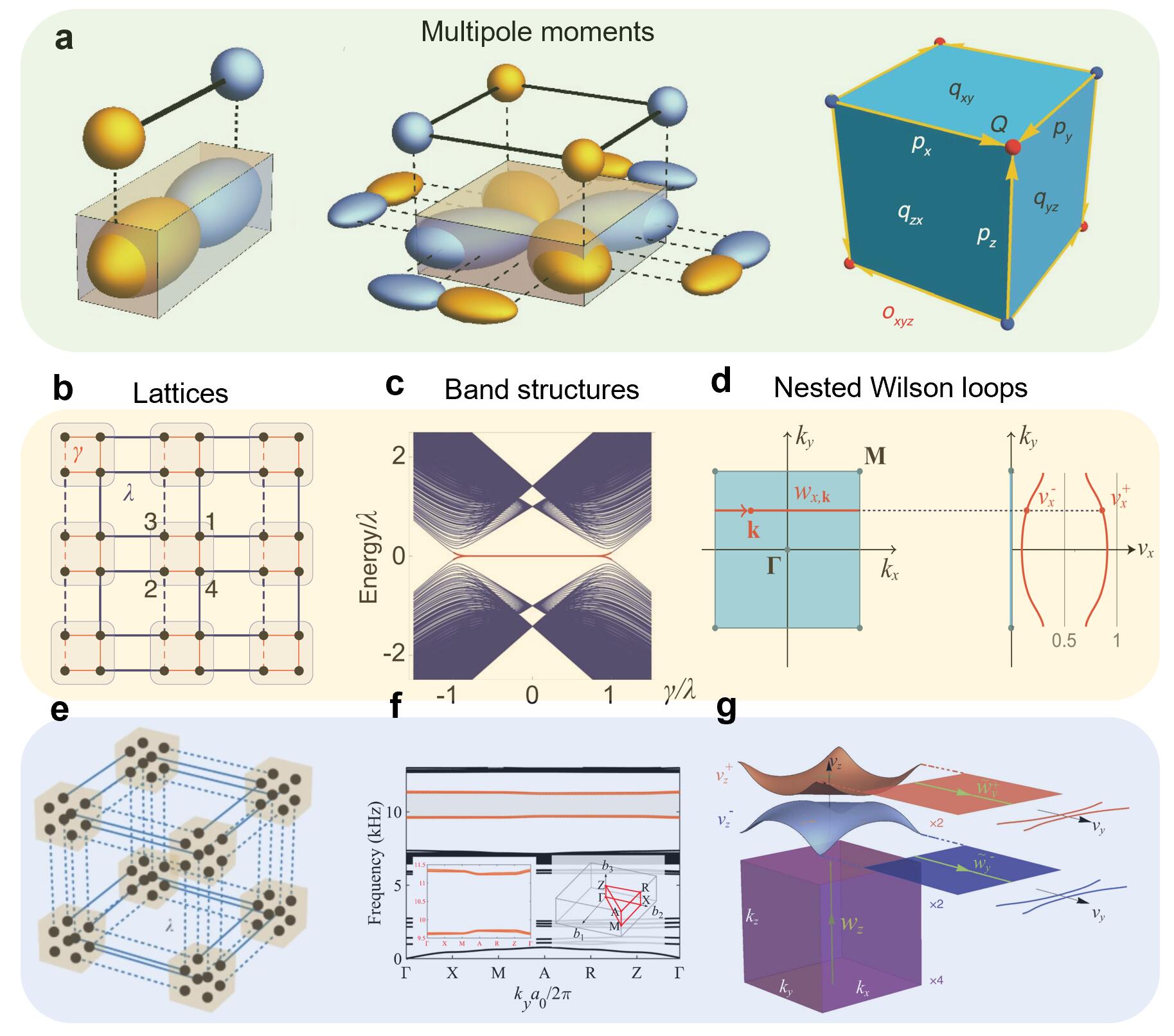}
\caption{\textbf{HOTIs with quantized multipole moments.} \textbf{a} Left panel: The bulk dipole moment induces edge charges. Middle panel: The bulk quadrupole moment induces edge dipole moments and corner charges. Right panel: The bulk octupole moment induces surface quadrupole moments, edge dipole moments and corner charges.\textbf{b} Schematics of the 2D tight-binding realizations of quadrupole insulators. Solid line represent positive hopping and dashed line represent negative hopping. \textbf{c} The evolution of the band gap and edge states regarding the change of quotient between intercell coupling and intracell coupling. \textbf{d} Schematics of the nested Wilson loop for a quadrupole insulator. \textbf{e} Schematics of the 3D tight-binding realizations of octupole insulators. Solid line represent positive hopping and dashed line represent negative hopping. \textbf{f} The band structure of of a phononic octupole insulator. \textbf{g} Schematics of the nested Wilson loop for an octupole insulator. Panels in \textbf{a} are reproduced from Ref.~\cite{HOTIEXP1,HOTI1}. Panels in \textbf{b-d, e, g} are reproduced from Ref.~\cite{HOTI1}. Panels in \textbf{c} are reproduced from Ref.~\cite{3D2}}.

\label{fig}
\end{figure}

\newpage

\begin{figure}[ht]
\centering
\includegraphics[width=\linewidth]{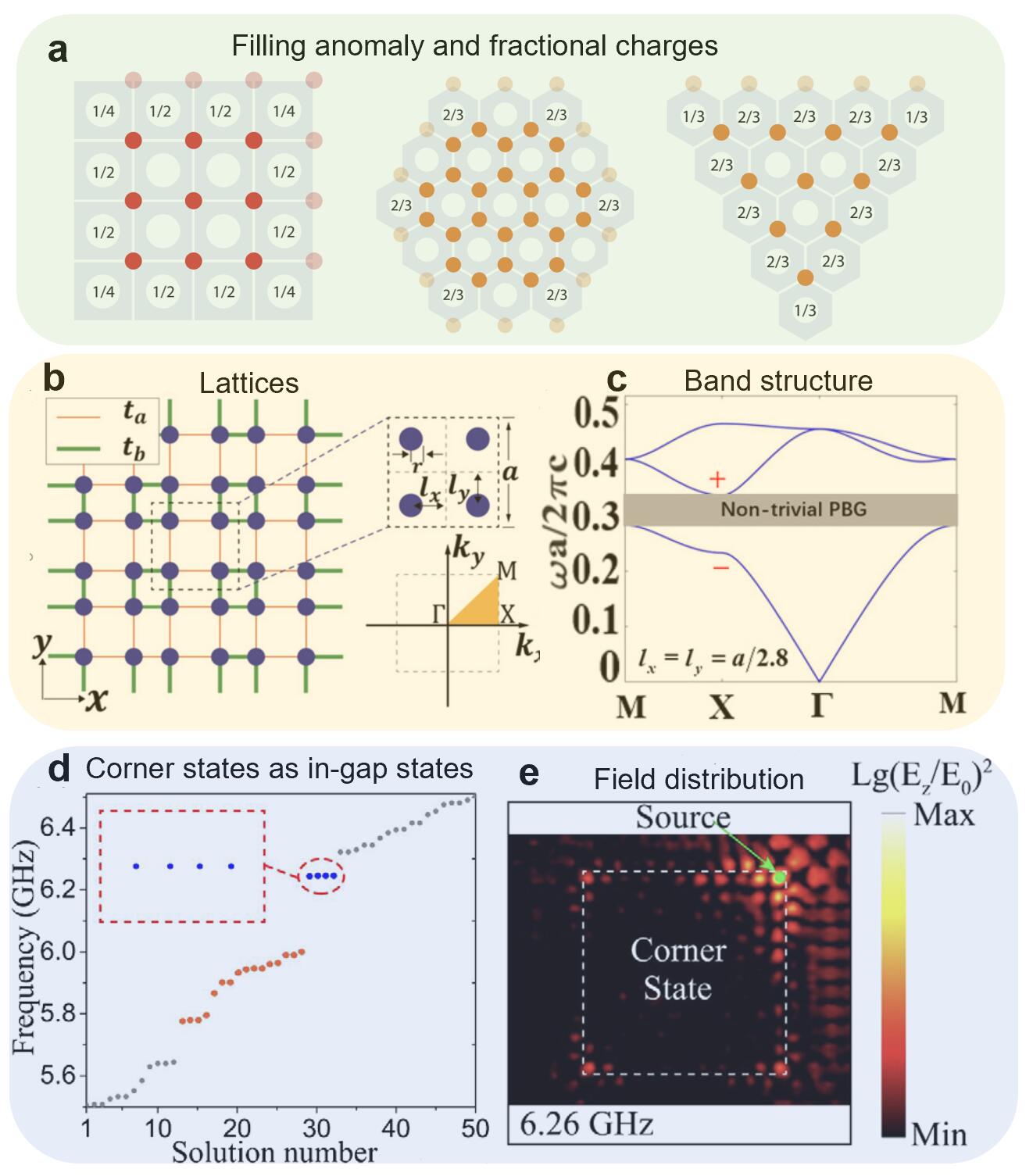}
\caption{\textbf{HOTIs without quantized multipole moments.} \textbf{a} Filling anomaly and fractional charges in 2D generalizations of SSH models with $C_4$ (Left panel), $C_6$ (Middle panel)and $C_3$ (Right panel). \textbf{b} Lattice structure of 2D photonic Su-Schrieffer-Heeger (SSH) models. \textbf{c} The band structure of 2D photonic SSH model in a topologically nontrivial phases. \textbf{d} The eigenvalues of a finite size 2D photonic SSH model. \textbf{e} The experimental excitation of corner states.Panels in \textbf{a} are reproduced from Ref.~\cite{HOTIC}. Panels in \textbf{b-c} are reproduced from Ref.~\cite{non1}. Panels in \textbf{d-e} are reproduced from Ref.~\cite{BIYEPRL}.}

\label{fig}
\end{figure}

\begin{figure}[ht]
\centering
\includegraphics[width=\linewidth]{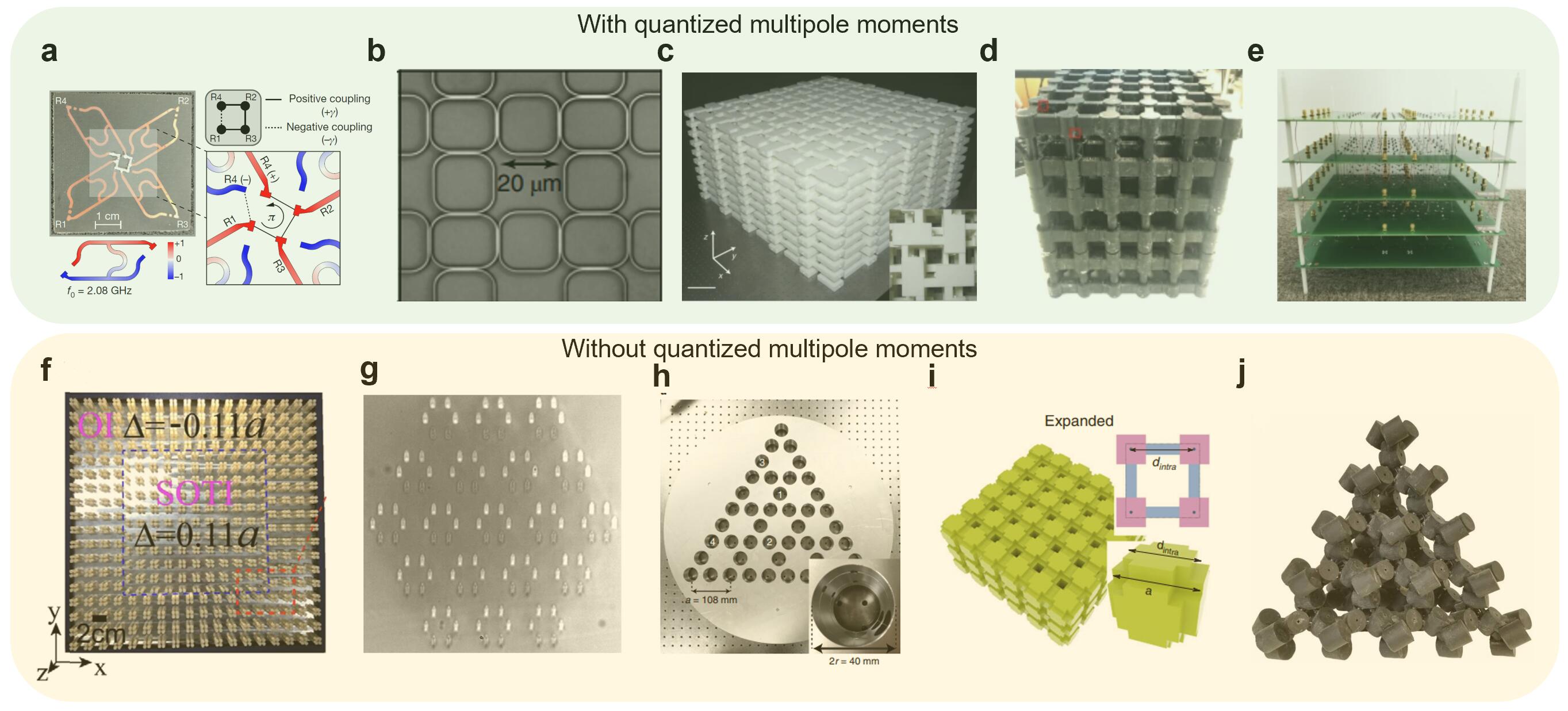}
\caption{\textbf{Various experimental realizations of HOTIs.}  Experimental realizations of quantized multipole insulator in \textbf{a} 2D microstrips, \textbf{b} 2D coupled resonantor optical waveguides,\textbf{c-d} 3D acoustic resonantors networks, \textbf{e} 3D electric circuits. The experimental photograph of \textbf{f} a 2D HOTIs with $C_4$ symmetry realized in all-dielectric photonic crystals, \textbf{g} a 2D HOTIs with $C_6$ symmetry in femto-second laser direct writing waveguides, \textbf{h} a 2D HOTIs with $C_3$ symmetry in acoustic resonator arrays, \textbf{i} a 3D HOTIs with cubic geometry in airborne sonic crystalsand \textbf{j} a 3D HOTI pyrochlore lattice in acoustic resonator arrays. Panels in \textbf{a}, \textbf{b}, \textbf{c}, \textbf{d}, \textbf{e}, \textbf{f}, \textbf{g}, \textbf{h}, \textbf{i}and \textbf{j} are reproduced from Ref.~\cite{HOTIEXP2}, Ref.~\cite{2019NP1}, Ref.~\cite{3D1}, Ref.~\cite{3D2}, Ref.~\cite{3D4}, Ref.~\cite{non1}, Ref.~\cite{non2}, Ref.~\cite{non3}, Ref.~\cite{HOTI3D2}and Ref.~\cite{HOTI3D3}, respectively.}
\label{fig}
\end{figure}

\newpage

\begin{table}[ht]
\centering
\begin{tabular}{p{170pt}p{50pt}p{120pt}p{110pt}}\toprule
\hline
\textbf{classical systems} & \textbf{Dimension of systems} & \textbf{Type of HOTIs} & \textbf{Uniqueness }\\
\hline
Dielectric photonic crystals~\cite{non1,BIYEPRL,DJWPRL,SB,2020NC,AS,HOTISPIN2} & 2 & Quadrupole insulators, generalized 2-dimensional (2D) Su-Schrieffer-Heeger (SSH) lattice, spinful 2D higher-order topological insulators (HOTIs)& scale invariance, low loss \\
\hline
Metallic photonic crystals~\cite{optica,nano} & 2 & Generalized 2D SSH lattice& Nano-fabrication, polarization dependent \\
\hline
Coupled resonators waveguide optical arrays~\cite{2019NP1} & 2 & Quadrupole insulators & Hopping phase controllable, optical frequency, compatible with quantum optics \\
\hline
Optical waveguides~\cite{non2,2019NP2,quantum} & $3$ & Generalized 2D SSH lattice & Time modulation, compatible with quantum optics, optical frequency \\
\hline
Airborne phononic crystals~\cite{HOTIEXP1,3D1,3D2,AM,non3,non4,zhang1,HOTI3D1,HOTI3D2,HOTI3D3,HOTI3D4,HOTI3D5,HOTIsyn3,nh10,HOTISPIN4} & 2 \& 3 & {Quadrupole insulators, Octupole insulators, Generalized 2D SSH lattices, 3D HOTIs, non-Hermitian 2D SSH lattice} & Direct mimicking tight-binding models, easy fabrications \\
\hline
Mechanics~\cite{HOTIEXP1,johan1,elastic,johan2,lamb} & 2 & Quadrupole insulators, Generalized 2D SSH lattices, Majorana corner states & scale invariance, high frequency \\
\hline
Electric circuits~\cite{HOTIEXP3,3D3,3D4,nonlinear3} & 2 \& 3 & Quadrupole insulators, Octupole insulators, nonlinear generalized 2D SSH lattices & Direct mimicking tight-binding models \\
\hline
Microstrips~\cite{HOTIEXP2,frac1,frac2} & 2 & Quadrupole insulators, Generalized 2D SSH lattices & Cheap, light and compact \\
\hline
Surface plasmonics~\cite{jiang3} & 2 & Quadrupole insulators & High frequency, subwavelength \\
\hline
\end{tabular}
\caption{\label{tab}\textbf{Summary of various realizations of HOTIs in classical waves.}}
\end{table}

\end{document}